\documentclass[letterpaper]{article} 
\usepackage{aaai2026}  
\usepackage{times}  
\usepackage{helvet}  
\usepackage{courier}  
\usepackage[hyphens]{url}  
\usepackage{graphicx} 
\usepackage{natbib}  
\usepackage{caption} 
\usepackage{algorithm}
\usepackage{algorithmic}
\usepackage{bm}
\usepackage{amsmath}
\usepackage{amssymb}
\usepackage{booktabs}
\usepackage{arydshln}
\usepackage{multirow}
\usepackage{newfloat}
\usepackage{listings}
\DeclareCaptionStyle{ruled}{labelfont=normalfont,labelsep=colon,strut=off} 
\floatstyle{ruled}
\newfloat{listing}{tb}{lst}{}
\floatname{listing}{Listing}
\title{Voices, Faces, and Feelings: Multi-modal Emotion-Cognition Captioning for Mental Health Understanding}
\author {
    Zhiyuan Zhou\textsuperscript{\rm 1},
    Yanrong Guo\textsuperscript{\rm 1}\thanks{Corresponding authors.},
    Shijie Hao\textsuperscript{\rm 1}\footnotemark[1]
}
\affiliations {
    \textsuperscript{\rm 1}Hefei University of Technology\\
    zhouzhiyuan.hfut@gmail.com, yrguo@hfut.edu.cn, hfut.hsj@gmail.com
}

\usepackage{bibentry}

\begin{document}

\maketitle

\begin{abstract}
Emotional and cognitive factors are essential for understanding mental health disorders. However, existing methods often treat multi-modal data as classification tasks, limiting interpretability especially for emotion and cognition. Although large language models (LLMs) offer opportunities for mental health analysis, they mainly rely on textual semantics and overlook fine-grained emotional and cognitive cues in multi-modal inputs. While some studies incorporate emotional features via transfer learning, their connection to mental health conditions remains implicit.
To address these issues, we propose ECMC, a novel task that aims at generating natural language descriptions of emotional and cognitive states from multi-modal data, and producing emotion–cognition profiles that improve both the accuracy and interpretability of mental health assessments. We adopt an encoder–decoder architecture, where modality-specific encoders extract features, which are fused by a dual-stream BridgeNet based on Q-former. Contrastive learning enhances the extraction of emotional and cognitive features. A LLaMA decoder then aligns these features with annotated captions to produce detailed descriptions.
Extensive objective and subjective evaluations demonstrate that: 1) ECMC outperforms existing multi-modal LLMs and mental health models in generating emotion–cognition captions; 
2) the generated emotion–cognition profiles significantly improve assistive diagnosis and interpretability in mental health analysis.
\end{abstract}


\section{Introduction}
Mental health disorders, such as depression, significantly affect physical and psychological well-being. According to the World Health Organization (WHO), depression affects over 300 million people worldwide, and untreated mental disorders will account for 13\% of the total disease burden by 2030 \cite{world2017depression}. One reason many patients struggle to receive evaluation is the lack of effective assistive mental health analysis methods. Current diagnostic methods rely heavily on self-reports and clinical interviews \cite{habtamu2023interventions}, which are subjective and inefficient. The shortage of psychiatrists cannot keep up with increasing patients \cite{johnson2022addressing}, highlighting the urgent need for efficient and scalable assistive mental health analysis.

\begin{figure}[htbp]  
    \centering
    \includegraphics[width=\linewidth]{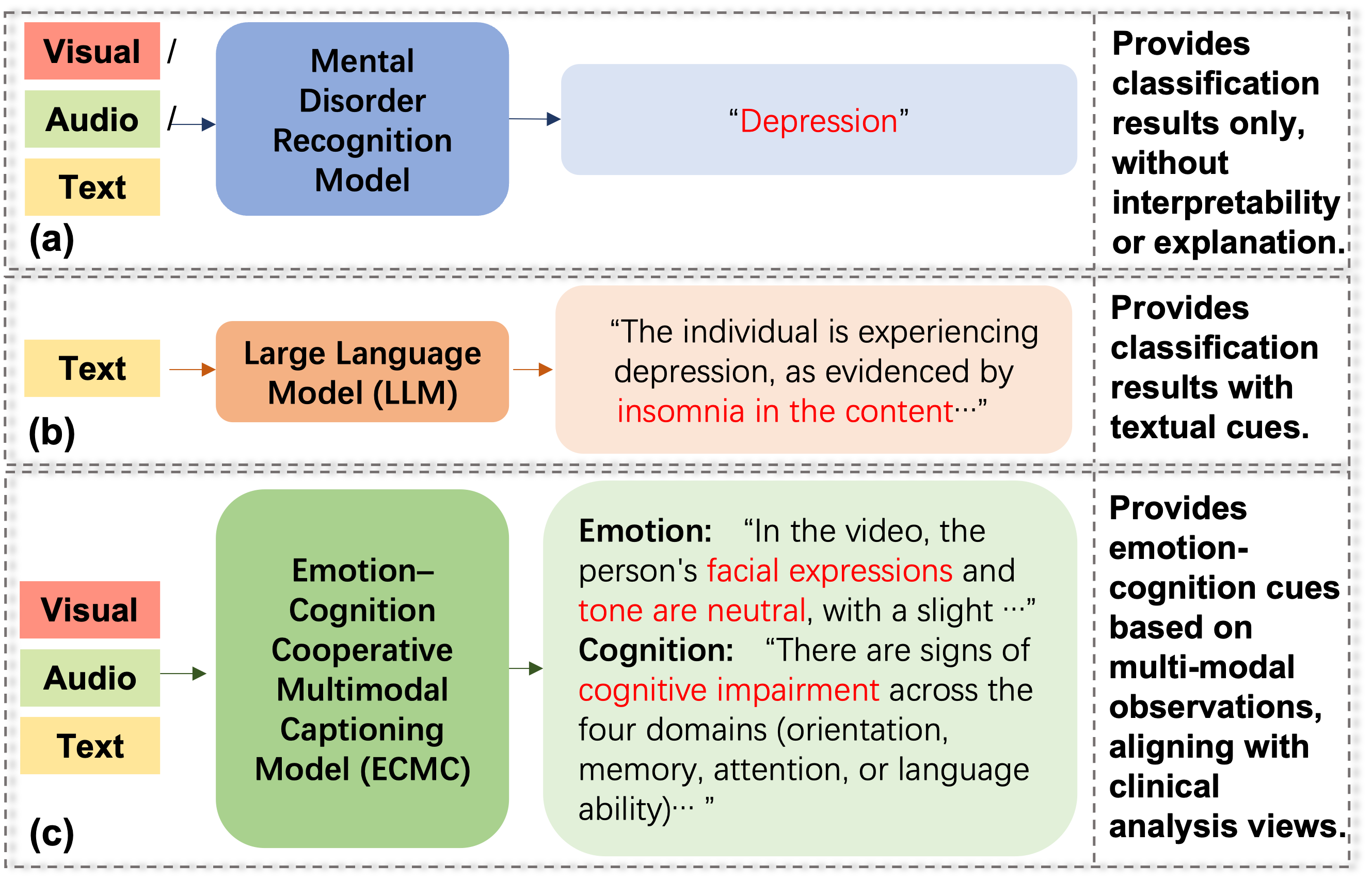}
    \caption{Different paradigms for mental disorder detection. (a) Classification-based. (b) LLM-based. (c) Ours.}
    \label{fig:1}
\end{figure}

However, existing assistive mental health analysis methods face challenges.
Firstly, most methods \cite{liu2024depression,ji2021mentalbert,zhang2022psychiatric} treat the task as a classification problem by categorizing multi-modal data to a mental disorder category, which lacks interpretability (Figure 1(a)). They do not reveal which cues are associated with mental health conditions. As a result, these methods are often unconvincing for clinical diagnostic support.
Secondly, large language models (LLMs) have strong natural language understanding abilities \cite{guo2025deepseek,achiam2023gpt}. However, when applied to mental health analysis, they rely on textual information, such as detecting symptom-related words \cite{yang2024mentallama,xu2023leveraging} (Figure 1(b)). They cannot capture emotional and cognitive signals from other modalities, like facial expressions or vocal tone, which are important in real-world clinical observations.
Despite the above two categories of methods contain works \cite{10094910,wu2022climate} that introduce affective features using transfer learning, the high-dimensional features offer limited insight into the correlation between emotion and mental disorders. They also ignore the role of cognitive factors in mental health analysis.

Emotion–cognition patterns play a vital role in understanding mental health disorders. Neuroscience research \cite{Shackman_Fox_Seminowicz_2015} suggests mental disorders are not only reflected in observable symptoms, but also in the dynamic mechanisms of emotional and cognitive processing. For example, depression is often associated with a prolonged emotional stagnation and suppressed cognitive functioning, while anxiety disorders exhibit rapid emotional and cognitive fluctuations, often manifesting as panic attacks.
Prior studies \cite{uban2021emotion,li2022clinical} have also identified significant differences in emotional and cognitive characteristics between individuals with and without mental health conditions based on linguistic features. However, existing methods rarely consider a multi-modal perspective to examine these patterns, and often fail to provide interpretable insights for clinical diagnosis.

To address these limitations, we introduce an emotion–cognition cooperative multi-modal captioning (ECMC) task, which aims to generate interpretable emotion–cognition profiles for mental health understanding (Figure 1(c)). Inspired by recent work on emotion captioning task \cite{xu2024secap,ye2024dual}, ECMC requires models to extract emotion- and cognition-relevant features from diverse modalities, and generate coherent, high-quality natural language descriptions. Unlike traditional approaches that compress complex signals to categorical labels, ECMC leverages natural language to unify multi-modal information into rich, contextualized narratives that capture both emotional states and cognitive impairments. This not only enhances interpretability for clinicians but also supports fine-grained comparisons across individuals and time, making ECMC a promising direction for accurate and explainable mental health assessment.

To this end, we design a multi-modal encoder–decoder framework, which compromises modality-specific encoders, a dual-stream BridgeNet, and a LLaMA-based decoder. Given the challenge of training encoders from scratch with limited data, we initialize the modality-specific encoders using pre-trained models tailored to each modality to extract initial features. However, such pre-trained models typically focus on frame-level representations and fail to capture emotion- and cognition-relevant semantics. Inspired by BLIP-2 \cite{li2023blip}, we design a dual-stream BridgeNet based on Q-former, which compresses multi-modal features and disentangles emotion and cognition representations with contrastive learning. Finally, we employ a LLaMA-based decoder, leveraging the powerful language understanding abilities of LLM, to generate emotion–cognition captions from the encoded features. The decoder guides the BridgeNet in the training process, to align emotion-cognition features with the target textual space.

The contributions of our work are summarized as follows:
\begin{itemize}
\item We propose a novel emotion–cognition cooperative multi-modal captioning (ECMC) task, which enables LLMs to generate emotion–cognition profiles for mental health understanding. To the best of our knowledge, this perspective has not been explored in previous work.
\item We design a multi-modal encoder–decoder framework for the ECMC\footnote{Code and more results: \url{https://github.com/zhouzyhfut/ECMC}.} task, which extracts emotion and cognition features from multi-modal data and aligns them with the LLM input space to produce informative captions.
\item Extensive objective and subjective evaluations demonstrate that our method can generate high-quality emotion–cognition captions from multi-modal observations.
\end{itemize}

\begin{figure*}[t]  
    \centering
    \includegraphics[width=\textwidth]{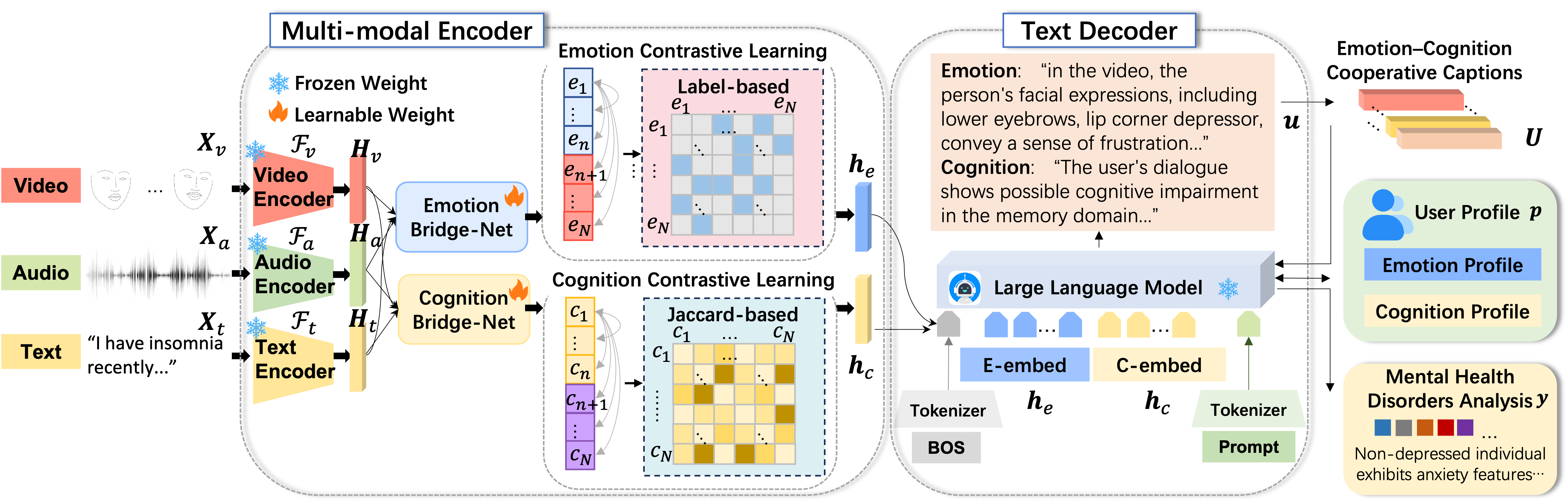}
    \caption{The framework of the proposed method.}
    \label{fig:2}
\end{figure*}

\section{Related Work}
\subsection{Mental Disorders Detection}
Most mental disorder detection methods focus on designing various deep learning models to classify samples into disorder categories \cite{liu2024depression,ji2021mentalbert,zhang2022psychiatric}. However, the underlying reasons behind these predictions are often implicit, limiting their utility in clinical applications.
The emergence of LLMs enhances mental disorders detection by fine-tuning on relevant data \cite{yang2024mentallama,xu2023leveraging}. Nevertheless, these approaches typically rely solely on textual semantic information and cannot capture facial expressions or other observable reactions from patients.
Although prior works \cite{uban2021emotion,li2022clinical} have incorporated affective information via transfer learning, the relationship between emotion-related features and prediction outcomes remains difficult to interpret.
In contrast, our approach enables multi-modal observation of patients’ emotion–cognition patterns and generates descriptions of salient cues, thus providing more interpretable evidence to support effective clinical diagnosis.

\subsection{Multi-modal Large Language Model}
Recent advances in multi-modal large language models (MLLMs) have achieved remarkable performance in cross-modal understanding and generation tasks.
For example, BLIP-2 \cite{li2023blip} introduces a lightweight query-based transformer that effectively bridges frozen vision encoders and language models, enabling strong zero-shot performance. 
Qwen2.5-Omni \cite{xu2025qwen2} supports seamless multi-turn dialogue grounded in visual content, highlighting its strength in multi-modal scenarios. 
InternVL-2.5 \cite{chen2024internvl} unifies visual and textual representations through dense alignment, enhancing complex cross-modal reasoning and fine-grained understanding. 
Sa2VA \cite{yuan2025sa2va} focuses on fine-grained semantic alignment, leading to  improvements in visual question answering and detailed image captioning.
Overall, MLLMs demonstrate impressive generalization across diverse modalities, laying a foundation for downstream multi-modal applications.

\subsection{Automatic Captioning}
Automatic captioning methods typically aim to generate natural language descriptions from visual or multi-modal inputs by leveraging encoder–decoder architectures.
For example, generating emotion–cause–aware video captions helps capture fine-grained multi-modal cues, which can then be used to summarize the causes of emotions in conversations \cite{wang2024observe}.
SECap \cite{xu2024secap} uses HuBERT and Q-Former to extract emotion-focused speech features and LLaMA to generate natural language captions of speech emotions.
By providing fine-grained and semantically rich textual descriptions, captioning tasks naturally offer detailed cues that can be highly valuable for mental health understanding, enabling interpretable insights into patients’ emotional and cognitive states.

\section{Method}
We first provide an overview of our ECMC. Then, we describe in detail how the emotion bridge-net and cognition bridge-net extract emotional and cognitive features, respectively. The overall training process is finally described. 

\subsection{Model Architecture}
As illustrated in Figure 2, inspired by automatic captioning in emotion recognition \cite{wang2024observe,xu2024secap,ye2024dual}, we adopt an encoder–decoder architecture.

\subsubsection{Encoder} 
For a given utterance sample $\bm{x}_i = \{\bm{X}_v, \bm{X}_a, \bm{X}_t\}$ where the subscripts $v$, $a$, and $t$ denote the video, audio, and text modalities respectively, we first extract modality-specific initial representations with modal-specific pre-trained models:
\begin{equation}
\bm{H}_v = \mathcal{F}_v(\bm{X}_v), \quad
\bm{H}_a = \mathcal{F}_a(\bm{X}_a), \quad
\bm{H}_t = \mathcal{F}_t(\bm{X}_t).
\end{equation}
Here, we employ VideoMAE \cite{tong2022videomae}, HuBERT \cite{soft-vc-2022}, and BERT \cite{devlin2019bert} as $\mathcal{F}_v$, $\mathcal{F}_a$, and $\mathcal{F}_t$, respectively.
 $\bm{H}_m \in \mathbb{R}^{T_m \times d_m}, m \in \{v,a,t\}$ represents initial representation of each modality, where $T_m$ is the number of time steps and $d_m$ is the dimension of representation.

Subsequently, we design a dual-stream BridgeNet comprising the emotion bridge-net and the cognition bridge-net, which fuses multi-modal information and further compresses the initial representations, since the sequential representations are often redundant. This process yields two embeddings, E-embedding and C-embedding, denoted as:
\begin{equation}
\{\bm{h}_e, \bm{h}_c\} 
= \mathcal{F}_{bridge}(\bm{H}_v, \bm{H}_a, \bm{H}_t)
\end{equation}
Here, $\bm{h}_e \in \mathbb{R}^{1\times d_e}$ captures features relevant to various emotional categories, while 
$\bm{h}_c \in \mathbb{R}^{1\times d_c}$ encodes features related to potential cognitive impairments.

\subsubsection{Decoder} 
We then employ LLaMA \cite{touvron2023llama} as a text decoder to generate fine-grained emotion–cognition descriptions. Leveraging the powerful language understanding and generation capabilities of LLMs, we guide the BridgeNet toward a semantically rich textual feature space and produce high-quality captions.
Specifically, we prepend a \texttt{<BOS>} token to the input, and concatenate it with the E-embedding $\bm{h}_e$ and C-embedding $\bm{h}_c$, followed by the tokenized prompt. The input to LLM is thus formulated as:
\begin{equation}
\bm{u} = \mathcal{F}_{llm}\bigl(concat(\texttt{<BOS>}, \bm{h}_e, \bm{h}_c, Prompt)\bigr)
\end{equation}
which constrains the output space of LLM to generate accurate captions $\bm{u}$ that reflect both emotional and cognitive aspects of the utterance.

After obtaining the emotion–cognition descriptions $\bm{U}=\{\bm{u}_1,...,\bm{u}_n\}$ for multiple utterances, we instruct LLM to produce a user profile $\bm{p}$. This profile summarizes salient patterns such as negative emotions and cognitive impairments observed in the utterances:
\begin{equation}
\bm{p} = \mathcal{F}_{llm}(\bm{U}, Prompt)
\end{equation}
The user profile $\bm{p}$ provides valuable evidence for mental disorder diagnosis, and further assists a general-purpose LLM in enhancing its mental disorder detection capabilities:
\begin{equation}
\bm{y} = \mathcal{F}_{llm}(\bm{p}, Prompt, \bm{X}_t)
\end{equation}
where $\bm{y}$ is the prediction of mental disorder.

\subsection{Emotional Representation Extraction}
To fuse multi-modal features and further compress the initial representations for extracting an emotional representation to feed into the decoder, we design the Emotion BridgeNet, as illustrated in Figure 3.

\subsubsection{Q-former}
For each modality $m$, we introduce a set of learnable query tokens $\bm{Q}_m \in \mathbb{R}^{l_q \times d_q}$, where $l_q$ is the number of query tokens and $d_q$ is the hidden dimension. We first apply self-attention over the query tokens to model intra-query dependencies:
\begin{equation}
\bm{Q}'_m = \mathrm{softmax} \left( \frac{\bm{Q}_m \bm{W}_q (\bm{Q}_m \bm{W}_k)^\top}{\sqrt{d_k}} \right) \bm{Q}_m \bm{W}_v
\end{equation}
where $\bm{W}_q \in \mathbb{R}^{d_q \times d_k},\bm{W}_k \in \mathbb{R}^{d_q \times d_k},\bm{W}_v \in \mathbb{R}^{d_q \times d_v}$ are the learnable weight matrices for queries, keys, and values.
Next, the updated queries $\bm{Q}'_m \in \mathbb{R}^{l_q \times d_v}$ used as queries for the cross-attention:
\begin{equation}
\bm{Z}_m = \mathrm{softmax}\!\left(
\frac{\bm{Q}'_m \bm{W}_q'^{(c)} (\bm{H}_m'\bm{W}_k'^{(c)})^\top}{\sqrt{d_k}}
\right)\bm{H}_m' \bm{W}_v'^{(c)}
\end{equation}
where $\bm{H}_m' = \bm{H}_m \bm{W}_m $, $\bm{W}_q'^{(c)} \in \mathbb{R}^{d_v \times d_k}$, $\bm{W}_k'^{(c)} \in \mathbb{R}^{d_q \times d_k}$, $\bm{W}_v'^{(c)} \in \mathbb{R}^{d_q \times d_v}$, $\bm{W}_m \in \mathbb{R}^{d_m \times d_q}$ are the learnable weight matrices.
Subsequently, the attended query representation $\bm{Z}_m \in \mathbb{R}^{l_q \times d_v}$ is passed through a feed-forward layer:
\begin{equation}
\bm{Z'}_m  = ReLU(\bm{Z}_m \bm{W}_1 + \bm{b}_1)\,\bm{W}_2 + \bm{b}_2
\end{equation}
where $\bm{W}_1 \in \mathbb{R}^{d_v \times d_e}$, $\bm{W}_2 \in \mathbb{R}^{d_e \times \frac{d_e}{3} }$, and $\bm{b}_1,\bm{b}_2$ are biases.

We concatenate the representations $\bm{Z'}_m $ and project them into the feature space:
\begin{equation}
\bm{h}_e = Norm(Pooling(Concat(\bm{Z'}_v, \bm{Z'}_a, \bm{Z'}_t) \, \bm{W}_3 + \bm{b}_3)), 
\end{equation}
where $\bm{h}_e  \in \mathbb{R}^{l_d \times d_e}$, $\bm{W}_3 \in \mathbb{R}^{d_e \times d_e}$ is the projection matrix and $\bm{b}_3$ is the bias. Pooling indicates meanpooling operation, and Norm is L2 normalization.

\begin{figure}[t]  
    \centering
    \includegraphics[width=\linewidth]{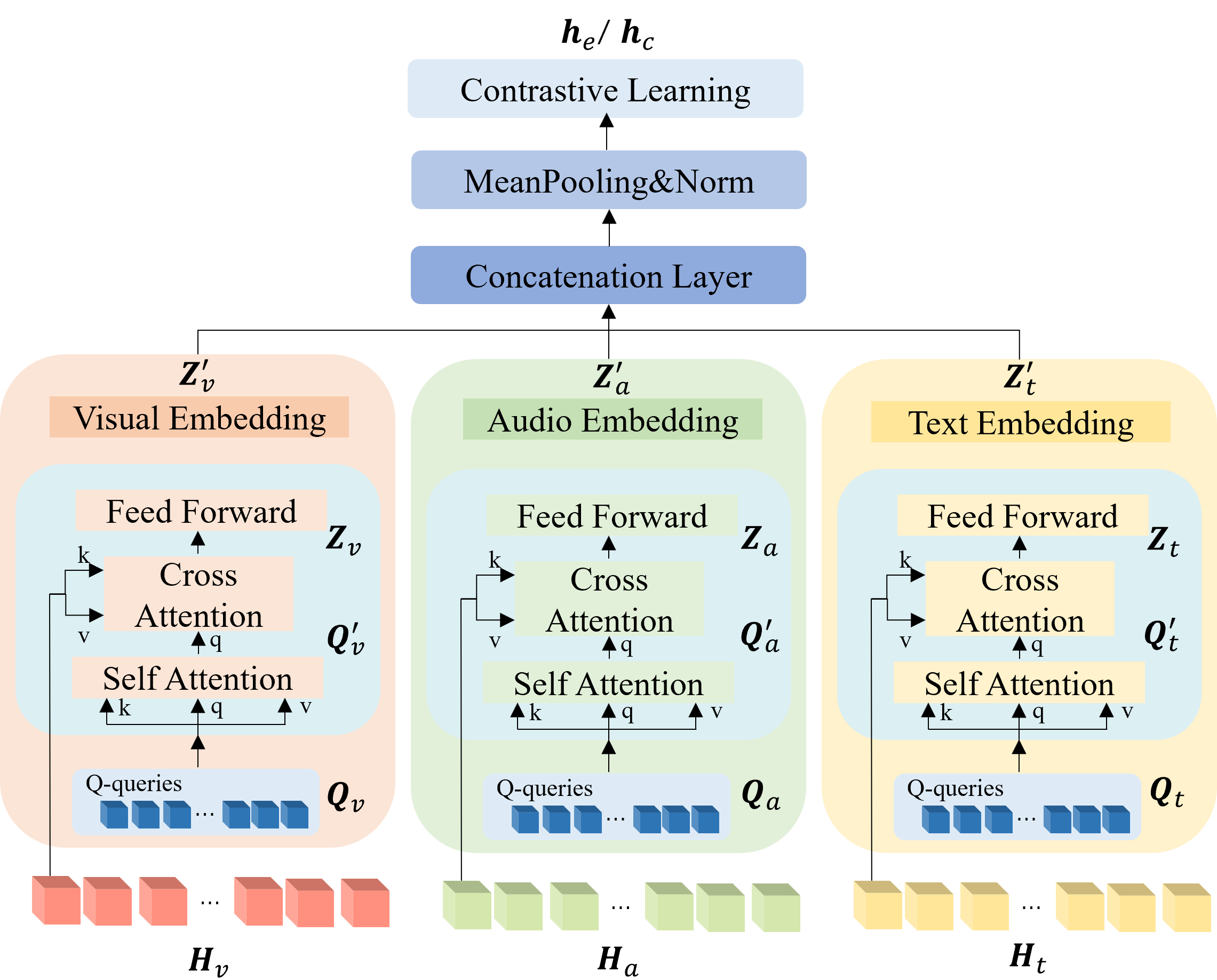}
    \caption{The architecture of BridgeNet.}
    \label{fig:3}
\end{figure}

\subsubsection{Emotion Contrastive Learning}
For mental disorder analysis, psychological researches \cite{joormann2016examining, vanderlind2020understanding,zakiei2014relationship} have highlighted negative emotions as a significant characteristic. Accordingly, we divide the feature representations $\bm{h}_{e,i}$ into negative, neutral, and positive categories according to the valence dimension \cite{kensinger2004remembering} in the affective space. To better capture the representations of different emotional categories, we introduce a label-matching based contrastive learning approach. We first compute the similarity matrix: $\bm{S}_{ij} = \frac{\bm{h}_{e,i}^\top \bm{h}_{e,j}}{\tau}.$ Then we build a binary mask $\bm{M}_{ij} = 1$ if $y_{e,i} = y_{e,j}$ and $i \neq j$, and $0$ otherwise, where $y_{e} \in \{-1,0,1\}$ denotes emotion class label, to identify intra-class positives. The loss combines intra-class compactness and inter-class separability:
\begin{equation}
\begin{aligned}
\mathcal{L}_{\mathrm{emo}} =
& - \frac{1}{N} \sum_{i=1}^N \frac{1}{|\{j:\bm{M}_{ij}=1\}|}
    \sum_{j:\bm{M}_{ij}=1} \log p_{ij} \\
& + \frac{1}{N} \sum_{i=1}^N 
    \log \bigl(1 + \sum_{j:\bm{M}_{ij}=0} {\mathrm{exp}}(\bm{S}_{ij})\bigr),
\end{aligned}
\end{equation}
where $\log p_{ij} = \bm{S}_{ij} - \log \sum_{k \neq i} {\mathrm{exp}}(\bm{S}_{ik})$, and $N$ represents the batch size.

\begin{table*}[htbp]\label{tbl1}
\centering
\setlength{\tabcolsep}{1.6mm} 
\begin{tabular}{l|c|c|ccccccc}
\specialrule{1.5pt}{0pt}{0pt}
\textbf{Method} & \textbf{Modality} & \textbf{Domain} & BLEU-1 & BLEU-2 & BLEU-4 & METEOR & ROUGE-L & CIDEr & F\_BERT \\
\hline
InternVL-2.5-8B         & VT  &  \multirow{4}{*}{General}   & 8.35 & 3.30 & 0.96 & 10.06 & 13.98 & 0.96 & 6.48 \\
DeepSeek-VL-2-small     & VT  &                            & 2.98 & 1.33 & 0.42 & 5.99 & 10.00 & 0.42 & 2.02 \\
Sa2VA-8B                & VT  &                            & 14.36 & 7.05 & 2.14 & \underline{15.35} & \underline{20.34} & 2.14 & \underline{12.28} \\
Qwen2.5-Omni-7B         & AVT &                             & 12.74 & 4.86 & 1.42 & 13.31 & 16.99 & 1.42 & 8.93 \\
\cdashline{1-10}
PsycoLLM                & T   & \multirow{5}{*}{Mental}     & 4.86 & 1.76 & 0.50 & 5.68 & 7.68 & 0.50 & 0.35 \\
MindChat                & T   &                            & 5.44 & 1.75 & 0.50 & 6.73 & 9.71 & 0.50 & 0.87 \\
EmoLLM                  & T   &                            & 8.50 & 3.50 & 1.19 & 10.08 & 14.25 & 1.19 & 6.25 \\
CPsyCoun                & T   &                            & \underline{17.44} & \underline{7.24} & \underline{2.33} & 15.07 & 18.90 & \underline{2.33} & 9.30 \\
\cdashline{1-10}
\textbf{Ours}           & AVT &                            & \textbf{34.76} & \textbf{19.40} & \textbf{8.28} & \textbf{29.47} & \textbf{24.91} & \textbf{8.28} & \textbf{27.13} \\
\specialrule{1.5pt}{0pt}{0pt}
\end{tabular}
\caption{Comparison of emotion captioning performance with other methods. A,V,T denote audio, video, and text, respectively. Bold indicates the best, while underline indicates the second-best.}
\end{table*}

\begin{table*}[htbp]\label{tbl2}
\centering
\setlength{\tabcolsep}{1.6mm}
\begin{tabular}{l|c|c|ccccccc}
\specialrule{1.5pt}{0pt}{0pt} 
\textbf{Method} & \textbf{Modality} & \textbf{Domain} & BLEU-1 & BLEU-2 & BLEU-4 & METEOR & ROUGE-L & CIDEr & F\_BERT \\
\hline
InternVL-2.5-8B       & VT  &   \multirow{4}{*}{General}  & 8.17  & 3.05 & 0.84 & 9.35  & 16.52 & 0.84 & 18.43 \\
DeepSeek-VL-2-small   & VT  &                          & 8.49  & 4.40 & 1.72 & 10.34 & 15.45 & 1.72 & 6.75 \\
Sa2VA-8B              & VT  &                          & 15.82 & 6.24 & 1.61 & 13.15 & 22.59 & 1.61 & 18.78 \\
Qwen2.5-Omni-7B        & AVT &  & 12.82 & 5.30 & 1.48 & 12.04 & 19.56 & 1.48 & \underline{20.03} \\
\cdashline{1-10}
PsycoLLM              & T   &  \multirow{4}{*}{Mental} & 6.22  & 3.19 & 1.30 & 10.59 & 10.48 & 1.30 & 6.60 \\
MindChat              & T   &                          & 6.81  & 3.02 & 1.11 & 8.33  & 11.74 & 1.11 & 5.41 \\
EmoLLM                & T   &                          & \underline{17.00} & \underline{6.66} & \underline{1.98} & 13.29 & 19.08 & \underline{1.98} & 14.99 \\
CPsyCoun              & T   &  & 9.32  & 3.89 & 1.06 & \underline{18.88} & \underline{23.07} & 1.06 & 4.47 \\
\cdashline{1-10}
\textbf{Ours}         & AVT &                 & \textbf{35.92} & \textbf{25.05} & \textbf{15.32} & \textbf{35.86} & \textbf{39.82} & \textbf{15.32} & \textbf{41.04} \\
\specialrule{1.5pt}{0pt}{0pt} 
\end{tabular}
\caption{Comparison of cognition captioning performance with other methods}
\end{table*}

\subsection{Cognitive Representation Extraction}
Similar to the emotional representation extraction process, we employ a Q-former to compress and fuse the multi-modal initial features $\bm{H}_m$, resulting in the cognition representation $\bm{h}_c$.
Guided by clinical cognitive scale MMSE \cite{arevalo2015mini}, we focus on four specific cognitive impairments: orientation deficit, attention deficit, memory deficit, and language disorder, and aim to discover representations corresponding to these distinct cognitive categories.
Since a sample $\bm{h}_{c,i}$ may be associated with multiple cognitive impairments simultaneously, we design a multi-label contrastive learning objective based on the Jaccard similarity to better capture the shared and unique aspects of these impairments.
For cognition features $\bm{h}_{c,i}$, each sample may have multiple labels $\bm{y}_{c,i} \in \{0, 1\}^4$. We replace the binary mask with a soft similarity weight based on Jaccard index:
\begin{equation}
\bm{W}_{ij} =
\begin{cases}
1, & \bm{y}_{c,i}= \bm{y}_{c,j} = \varnothing \\
0, & \bm{y}_{c,i}= \varnothing \ \text{or}\ \bm{y}_{c,j} = \varnothing \\
\dfrac{|\bm{y}_{c,i}\cap \bm{y}_{c,j}|}{|\bm{y}_{c,i}\cup \bm{y}_{c,j}|}, & \text{otherwise}.
\end{cases}
\end{equation}
The loss softly pulls samples with higher label overlap closer and pushes others apart:
\begin{equation}
\begin{aligned}
\mathcal{L}_{\mathrm{cog}} =
- \frac{1}{N} \sum_{i=1}^N \log \frac{\sum_{j} {\mathrm{exp}}(\bm{S}_{ij}) \cdot \bm{W}_{ij}}{\sum_{j} {\mathrm{exp}}(\bm{S}_{ij})} \\
+ \frac{1}{N} \sum_{i=1}^N \log \bigl(1 + \sum_{j} {\mathrm{exp}}(\bm{S}_{ij}) (1 - \bm{W}_{ij})\bigr).
\end{aligned}
\end{equation}

\subsection{Training Process}
To enhance multi-modal emotion–cognition captions with LLMs, we design a two-stage training process. In the first stage, we jointly train the emotional representation extraction and cognitive representation extraction processes. This collaborative training effectively compresses and fuses multi-modal information, enabling the extraction of compact emotional and cognitive representations:
\begin{equation}
\mathcal{L}_1=\mathcal{L}_{\mathrm{emo}}+\mathcal{L}_{\mathrm{cog}},
\end{equation}
where we keep the parameters of the modal-specific encoder frozen, and the Q-Former is initialized with BERT \cite{devlin2019bert} pre-trained parameters.

In the second stage, we fine-tune the emotion Bridge-Net and cognition Bridge-Net to effectively integrate the extracted features into the LLM. Specifically, we keep both the modal-specific encoders and LLM parameters frozen, and employ cross-entropy loss to make the LLM generate annotated captions $\hat{\bm{u}}$,  thereby guiding the Bridge-Net to produce features aligned with the LLM's input space:
\begin{equation}
\mathcal{L}_2=CELoss(\hat{\bm{u}},\bm{u})
\end{equation}

\begin{table*}[htbp]
\centering
\setlength{\tabcolsep}{0.5mm}
\begin{tabular}{lcc|lcc|lcc|cc}
\specialrule{1.5pt}{0pt}{0pt}
\textbf{Method} & \textbf{ACC} & \textbf{F1} & 
\textbf{Method} & \textbf{ACC} & \textbf{F1} & 
\textbf{Method} & \textbf{ACC} & \textbf{F1} &
\textbf{Im\_ACC} & \textbf{Im\_F1} \\
\hline
Llama-3-8B &  44.95 &  46.43 & Qwen3-8B &  50.46 &  60.29 & GLM4-9B &  47.71 &  58.39 & 0.00 & 0.00\\
\cdashline{1-11}
InternVL-2.5-8B
    &  48.62 &  46.15 
    & InternVL-2.5-8B &  \underline{69.72} &  \underline{69.81} 
    & InternVL-2.5-8B &  53.21 &  59.84 & 9.48↑ &3.56↑ \\
DeepSeek-VL-2-small 
    &  41.28 &  47.54 
    & DeepSeek-VL-2-small  &  51.38 &  59.54 
    & DeepSeek-VL-2-small  &  51.38 &  60.15 & 0.31↑ & 0.71↑\\
Sa2VA-8B 
    &  48.62 &  49.09 
    & Sa2VA-8B  &  67.89 &  69.03 
    & Sa2VA-8B  &  \underline{57.80} &  62.90 & \underline{10.40↑} &5.30↑ \\
Qwen2.5-Omni-7B 
    &  \underline{49.54} &  54.24 
    & Qwen2.5-Omni-7B &  65.14 &  67.24 
    & Qwen2.5-Omni-7B &  56.88 &  58.46 & 9.48↑ &4.94↑\\
\cdashline{1-11}
PsycoLLM  
    &  47.71 &  44.66 
    & PsycoLLM &  54.13 &  62.12 
    & PsycoLLM &  53.21 &  61.07 & 3.98↑ &0.91↑ \\
MindChat 
    &  29.36 &  36.36 
    & MindChat &  58.72 &  62.81 
    & MindChat &  \underline{57.80} &  62.90 & 0.92↑ &1.01↓\\
EmoLLM
    &  45.87 &  \underline{54.96} 
    & EmoLLM &  61.47 &  63.16 
    & EmoLLM &  55.05 &  \underline{63.16} & 6.42↑& \underline{5.39↑} \\
CPsyCounX 
    &  48.62 &  48.15 
    & CPsyCounX &  67.89 &  68.47 
    & CPsyCounX &  55.05 &  60.80 & 9.48↑& 4.10↑ \\
\cdashline{1-11}
\textbf{Ours} 
    &  \textbf{50.46} &  \textbf{56.69} 
    & \textbf{Ours} &  \textbf{70.64} &  \textbf{70.27} 
    & \textbf{Ours} &  \textbf{59.63} &  \textbf{65.62} & \textbf{12.54↑} & \textbf{9.16↑}\\
\specialrule{1.5pt}{0pt}{0pt}
\end{tabular}
\caption{Depression Detection by Baseline LLMs Using Emotion–Cognition Profiles from Different Methods. Row 1: results of baseline models. 'Im\_ACC' and 'Im\_F1' show average gains over Llama-3-8B, Qwen3-8B, and GLM4-9B in ACC and F1.}
\end{table*}

\begin{table*}[htbp]
\centering
\setlength{\tabcolsep}{0.5mm}
\begin{tabular}{lcc|lcc|lcc|cc}
\specialrule{1.5pt}{0pt}{0pt}
\textbf{Method} & \textbf{ACC} & \textbf{F1} & 
\textbf{Method} & \textbf{ACC} & \textbf{F1} & 
\textbf{Method} & \textbf{ACC} & \textbf{F1} &
\textbf{Im\_ACC} & \textbf{Im\_F1}\\
\hline
Llama-3-8B &  38.54 &  52.80 
& Qwen3-8B &  41.67 &  57.58 
& GLM4-9B &  43.75 &  57.14 & 0.00 & 0.00\\
\cdashline{1-11}
InternVL-2.5-8B 
    &  \underline{56.25} &  56.25 
    & InternVL-2.5-8B &  55.21 &  63.87 
    & InternVL-2.5-8B &  42.71 &  58.02 & 10.07↑ & 3.54↑\\
DeepSeek-VL-2-small 
    &  48.96 &  55.05 
    & DeepSeek-VL-2-small  &  45.83 &  59.38 
    & DeepSeek-VL-2-small  &  41.67 &  57.58 & 4.17↑ & 1.50↑ \\
Sa2VA-8B 
    &  55.21 &  52.75 
    & Sa2VA-8B  &  54.17 &  62.71 
    & Sa2VA-8B  &  \underline{44.79} &  \underline{58.46} & 10.07↑ & 2.13↑\\
Qwen2.5-Omni-7B 
    &  50.00 &  50.00 
    & Qwen2.5-Omni-7B &  \underline{65.62} &  65.96 
    & Qwen2.5-Omni-7B &  43.75 &  \underline{58.46} & 11.80↑ & 2.30↑ \\
\cdashline{1-11}
PsycoLLM 
    &  52.08 &  \underline{57.41} 
    & PsycoLLM &  48.96 &  59.50 
    & PsycoLLM &  43.75 &  \underline{58.46} & 6.94↑& 2.62↑ \\
MindChat 
    &  47.92 &  46.81 
    & MindChat &  46.88 &  59.84 
    & MindChat &  41.67 &  57.58 & 4.17↑ & 1.10↓ \\
EmoLLM 
    &  54.17 &  55.10 
    & EmoLLM &  61.46 &  \underline{66.06} 
    & EmoLLM &  43.75 &  \underline{58.46} & \underline{11.81↑} &\underline{4.03↑}\\
CPsyCounX 
    &  46.88 &  54.05 
    & CPsyCounX &  56.25 &  63.16 
    & CPsyCounX &  42.71 &  58.02 & 7.29↑ & 2.57↑\\
\cdashline{1-11}
\textbf{Ours} 
    &  \textbf{57.29} &  \textbf{58.59} 
    & \textbf{Ours} &  \textbf{66.67} &  \textbf{69.16} 
    & \textbf{Ours} &  \textbf{44.95} &  \textbf{58.91} & \textbf{14.98↑} & \textbf{6.38↑}\\
\specialrule{1.5pt}{0pt}{0pt}
\end{tabular}
\caption{Anxiety Detection by Baseline LLMs Using Emotion–Cognition Profiles from Different Methods}
\end{table*}

\section{Dataset and Evaluation Metric}
\subsubsection{Dataset}
We evaluate our method on the MMDA dataset \cite{jiang2022mmda}, a large-scale multimodal benchmark for mental disorder analysis containing 1,025 clinically validated samples. Each sample includes visual, acoustic, and textual data from clinical interviews, with depression and anxiety labels based on standardized assessments (HAMD and HAMA).
To obtain utterance-level emotion–cognition captions, we combine automatic annotation with manual refinement. The interviews are first segmented into question–answer turns. Emotion-LLaMA \cite{cheng2024emotion} then generates emotional descriptions integrating facial, vocal, and semantic cues. Using these outputs and the textual content, DeepSeek-671B \cite{guo2025deepseek} produces corresponding cognitive captions and categorical labels for both emotion and cognition, yielding 30,592 annotated utterance–caption pairs.
To ensure data quality, non-informative segments (e.g., greetings, personal information inquiries, side conversations, or utterances shorter than 2 seconds) were removed to retain core doctor–patient interactions. Annotators further discarded inconsistent labels to minimize noise and bias. We computed inter-rater agreement, yielding substantial consistency (Fleiss’s $\kappa$ = 0.64 for emotion and 0.73 for cognition). After filtering and verification, the dataset comprises 13,536 segments for training, 1,402 for validation, and 3,790 for testing. The dataset’s category distribution is 0.91\% orientation deficit, 8.54\% attention deficit, 10.38\% memory deficit, 15.89\% language disorder, and 31.85\% negative emotion.

\textbf{Objective metrics} For the ECMC task, we adopt the natural language evaluation metrics as used in automatic captioning, including BLEU-1, BLEU-2, BLEU-4, METEOR, ROUGE-L, CIDEr, and F\_BERT. Among these, BLEU, METEOR, ROUGE-L, and CIDEr are primarily word-level matching metrics, while F\_BERT is a semantic similarity-based metric. For evaluating the generated emotion-cognition profiles on mental disorder detection, we use accuracy (ACC) and F1 score (F1). 

\textbf{Subjective Metrics} To evaluate both the quality of our dataset annotations and the effectiveness of captions generated by different methods, we conduct a human evaluation using a five-point Likert scale \cite{joshi2015likert} across the following designed dimensions: (1) Emotional Accuracy (EAcc), measuring whether the generated emotional captions accurately reflect the content of the video; (2) Cognitive Accuracy (CAcc), measuring whether the captions of cognitive impairments correctly correspond to the dialogue content; (3) Detail Richness (Detail), evaluating the richness and granularity of details in the captions; (4) Label Consistency (Consis), assessing whether the captions align with the corresponding classification labels; and (5) Overall Quality (Overall), providing an overall assessment of the  quality. 

\section{Experiments and Analysis}
\subsection{Experiment Setup}
For the first stage, we train our model on two NVIDIA RTX A6000 GPUs with a batch size of 64 for up to 500 epochs. The maximum sequence lengths of 1024 for audio and 512 for video. We use the AdamW optimizer with a learning rate of $1.3 \times 10^{-5}$, and a weight decay of $1.0 \times 10^{-6}$. The model has approximately 701M total parameters, with about 598M parameters being trainable. 
For the second stage, we train our model with a micro batch size of 8 per GPU and accumulation over 8 steps. We also apply DeepSpeed ZeRO stage-2 optimization and set mixed-precision training (fp16). In this stage, the model has approximately 7.6B total parameters, with about 605M trainable parameters.

\subsection{Performance Analysis}
We evaluate our method through caption comparison with baselines, mental disorder detection performance, ablations on modalities and contrastive modules, statistical analysis across disorders, and human evaluation.

\subsubsection{Comparison of Emotion–Cognition Captions}
We select general multi-modal large language models (MLLMs) (InternVL-2.5 \cite{chen2024internvl}, DeepSeek-VL-2 \cite{lu2024deepseek}, Sa2VA \cite{yuan2025sa2va}, Qwen2.5-Omni \cite{xu2025qwen2}), as well as large language models (LLMs) specifically fine-tuned for the mental health domain (PsycoLLM \cite{hu2024psycollm}, MindChat\footnote{\url{https://github.com/X-D-Lab/MindChat}}, EmoLLM\footnote{\url{https://github.com/SmartFlowAI/EmoLLM}}, CPsyCoun \cite{zhang2024cpsycoun}), for comparison. 
Tables 1 and 2 present the ECMC performance of different models. 
We can see that our model consistently outperforms other competitors in both emotion and cognition captioning. General-domain MLLMs like Sa2VA and Qwen2.5-Omni perform relatively well due to better instruction-following and broader data coverage, but often produce short or inconsistent outputs. Mental health LLMs, though domain-specific, are limited to text and struggle with describing non-verbal cues, leading to lower scores. In contrast, our method effectively integrates multi-modal data, including facial expressions, vocal tone, and text semantics, to generate more comprehensive and higher-quality emotion-cognition captions.

\begin{figure}[t]  
    \centering
    \includegraphics[width=\linewidth]{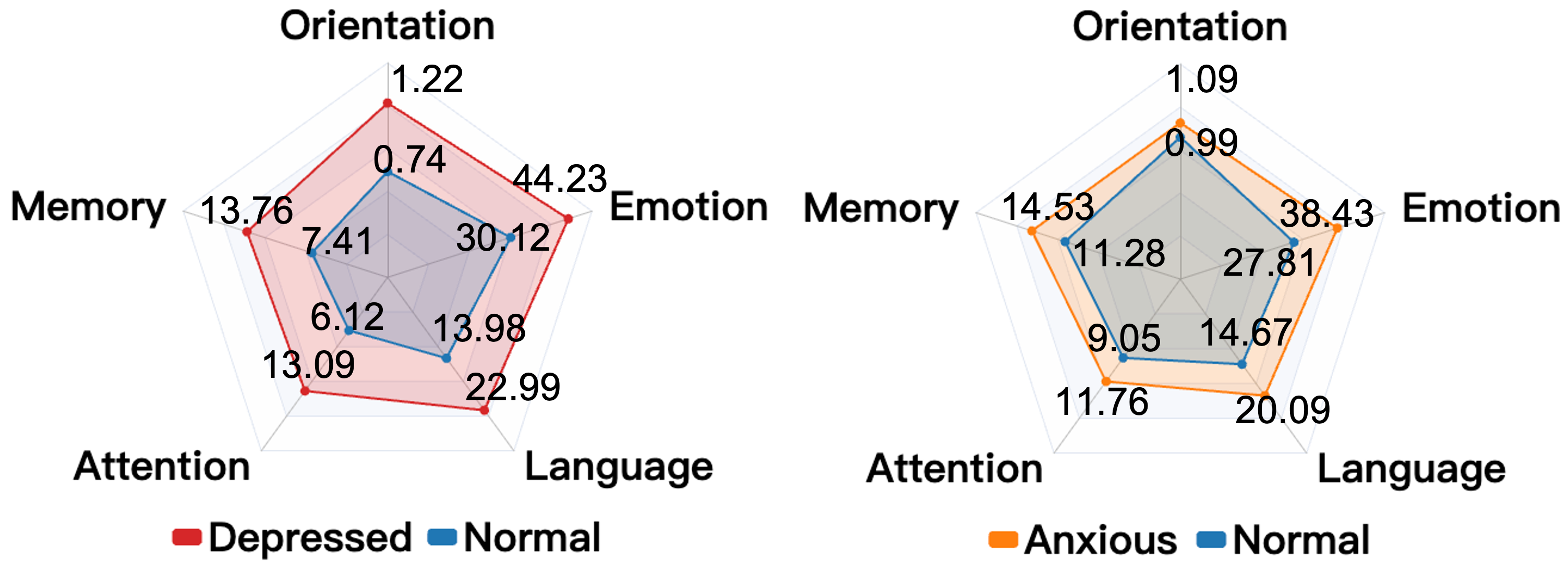}
    \caption{Emotional and cognitive differences between normal individuals and patients with different mental disorders}
    \label{fig:4}
\end{figure}

\begin{table}[htbp]
\centering
\setlength{\tabcolsep}{2mm}
\begin{tabular}{ccccc|c}
\specialrule{1.5pt}{0pt}{0pt}
\textbf{Audio} & \textbf{Video} & \textbf{Text} & \textbf{EmoCL} & \textbf{CogCL} & \textbf{F\_BERT} \\
\hline
\checkmark &  & & \checkmark & \checkmark & 18.10 \\
&  \checkmark &  & \checkmark & \checkmark &  16.30\\
& & \checkmark & \checkmark & \checkmark & 16.75 \\
\checkmark & \checkmark &           & \checkmark & \checkmark& 22.05 \\
\checkmark &  & \checkmark & \checkmark & \checkmark &  29.30 \\
&  \checkmark & \checkmark & \checkmark & \checkmark &  27.21 \\
\cdashline{1-6}
\checkmark & \checkmark & \checkmark &           &           & 23.89 \\
\checkmark & \checkmark & \checkmark &           &\checkmark & 27.24 \\
\checkmark & \checkmark & \checkmark & \checkmark &           & 26.94\\
\cdashline{1-6}
\checkmark & \checkmark & \checkmark & \checkmark & \checkmark & 34.09  \\
\specialrule{1.5pt}{0pt}{0pt}
\end{tabular}
\caption{Ablation Study on Different Modules. EmoCL and CogCL represent emotion contrastive learning and cognition contrastive learning, respectively.}
\label{ablation}
\end{table}

\subsubsection{Effect on Mental Disorder Detection}
We evaluate the impact of emotion–cognition profiles generated by different methods on depression and anxiety detection using existing baseline LLMs, including LLaMA-3-8B \cite{dubey2024llama}, Qwen3-8B \cite{yang2025qwen3}, and GLM4-9B \cite{glm2024chatglm}. For each subject, we first summarize the captions generated by different methods (Table 1 and 2) into emotion–cognition profiles using DeepSeek-671B. These profiles, together with the original dialogue content, are then provided as input to the baseline LLMs.
As shown in Table 3 and 4, the profiles generated by our method yield higher accuracy and F1 scores across both disorder types, significantly improving the performance of different baselines, with average ACC and F1 improvements over 12.54\% and 9.16\% on depression detection, and 14.98\% and 6.38\% on anxiety detection. Notably, we also observe that ineffective captions can even degrade detection performance, as long but uninformative texts may increase the difficulty for LLMs to extract relevant indicators. The results validate that our method produces higher-quality emotion–cognition profiles, which effectively enhance the accuracy of mental disorder detection using existing general models.

\subsubsection{Ablation Study}
We conduct ablation studies to verify the contribution of different modalities and contrastive learning modules to the final performance. We use the semantic similarity metric F\_BERT to evaluate the overall effectiveness of the emotion–cognition captions. As shown in Table 5,  among uni-modal inputs, audio achieves the best performance, indicating its discriminative power. Multi-modal combinations further improve results, with audio+text showing strong complementarity. Removing either EmoCL or CogCL causes the model to degenerate into a single-branch structure, resulting in performance drops, while removing both leads to a significant decline. This demonstrates the importance of the dual-branch structure for effectively decoupling emotion and cognition features. The full model with all modalities and contrastive objectives achieves the best score, confirming the effectiveness of each component.

\begin{table}[htbp]
\centering
\label{tab:Human}
\setlength{\tabcolsep}{1.2mm}
\begin{tabular}{l|ccccc}
\specialrule{1.5pt}{0pt}{0pt}
\textbf{Method/Dataset} & \textbf{EAcc} & \textbf{CAcc} & \textbf{Detail} & \textbf{Consis} & \textbf{Overall} \\
\hline
Dataset     &4.2 &  3.9 &      4.6 &     4.3  &       4.3          \\
\cdashline{1-6}
Sa2VA   & \underline{3.76}  &  4.40 &  \textbf{4.36}&   - & \underline{4.15} \\
Qwen2.5-Omni  & 3.73	&\underline{4.45}	&\underline{4.32}	& - &	4.12\\
EmoLLM      &3.6&	3.91&	3.96	& - &	3.79\\
CPsyCoun    &3.2 &	3.6	 &3.68 & - &		3.46 \\
\textbf{Ours}  & \textbf{3.98}	&\textbf{4.51}&	\underline{4.32}  & -  & \textbf{4.26}\\
\specialrule{1.5pt}{0pt}{0pt}
\end{tabular}
\caption{Human Evaluation on Dataset and Methods}
\end{table}

\subsubsection{Emotional and Cognitive Statistics across Different Mental Disorders}
To explore the differences in emotion–cognition patterns across mental disorders and healthy individuals, we calculate the proportion of negative emotions, orientation deficit, attention deficit, memory deficit, and language disorder in utterances of each individual, and then compute the average within each group (depression, anxiety, and healthy controls). As shown in Figure 4,  both depressed and anxious individuals exhibit significantly higher proportions of negative emotions compared to healthy participants. In terms of cognitive impairments, depression is associated with notably higher frequencies than anxiety. Among the cognitive categories, orientation deficits appear less frequently, while attention deficits, memory deficits, and language disorders are more prevalent. The above observations validate the necessity of the ECMC task.

\subsubsection{Human Evaluation}
We invited eight psychology practitioners with expertise in mental health to evaluate our dataset and model outputs. As shown in Table 6, the dataset receives an overall score of 4.3, indicating high-quality emotion-cognition captions. EAcc, Detail, and Consis all score above 4.0, suggesting accurate emotional content, rich detail, and reliable labels for contrastive learning. CAcc scores slightly lower, reflecting the difficulty of generating cognitively accurate descriptions from multi-modal and textual inputs, and highlighting the need for better multi-modal cognitive modeling.
We compare our method with two top-performing general MLLMs and two leading mental health LLMs. Our model achieves the highest scores in EAcc, CAcc, and Overall, and ranks second in Detail. Notably, MLLMs receive higher subjective ratings than mental health LLMs, empirically validating the effectiveness of combining visual and acoustic cues, which mental health LLMs lack.

\section{Conclusion}
In this paper, we introduce the ECMC task to generate interpretable emotion–cognition profiles from multi-modal data for mental health understanding. To support this task, we design a novel encoder–decoder framework that extracts and aligns emotion–cognition features with a LLaMA-based decoder. Experimental results show that our method produces high-quality, explainable captions, offering a promising direction for interpretable and fine-grained mental health assessment. 

\section{Acknowledgments}
This work was supported by the National Key Research and Development Program (Grant No. 2023YFC2506800), the National Natural Science Foundation of China under Grant No. 62072152, 62172137.

\nobibliography*

\bibliography{aaai2026}
\end{document}